\documentclass{article}
\usepackage{cite}
\usepackage{amsmath}

\reversemarginpar

\newcommand{\ret}{\hfill\break\noindent}
\newcommand{\nn}{\nonumber}

\newcommand{\ba}{\begin{eqnarray}}
\newcommand{\ea}{\end{eqnarray}}

\newcommand{\lb}{\ell_B}
\newcommand{\lbsq}{\ell_B^2}
\def\ie{{\it i.e.\ }}

\begin{document}
\begin{flushright} Preprint USM-TH-125 \end{flushright}
\medskip

\begin{center} \Large\bf Symmetry in noncommutative quantum mechanics\end{center}
\vspace*{.15in} \normalsize
\begin{center}
Olivier Espinosa\footnote{olivier.espinosa@fis.utfsm.cl}
and
Patricio Gaete\footnote{patricio.gaete@fis.utfsm.cl}
\end{center}

\vspace*{.15in}
\begin{center}
Departamento de F\'{\i}sica \\ Universidad T\'{e}cnica Federico
Santa Mar\'{\i}a \\ Casilla 110-V \\ Valpara\'{\i}so, Chile
\end{center}

\begin{abstract}
%
We reconsider the generalization of standard quantum mechanics in
which the position operators do not commute. We argue that the
standard formalism found in the literature leads to theories that
do not share the symmetries present in the corresponding
commutative system. We propose a general prescription to specify a
Hamiltonian in the noncommutative theory that preserves the
existing symmetries. We show that it is always possible to choose
this Hamiltonian in such a way that the energy spectrum of the
standard and non-commuting theories are identical, so that
experimental differences between the predictions of both theories
are to be found only at the level of the detailed structure of the
energy eigenstates.
\end{abstract}

\vskip 2cm
\noindent
Corresponding Author:\\
Olivier Espinosa (olivier.espinosa@fis.utfsm.cl)\\
Departamento de F\'{\i}sica, Universidad T\'{e}cnica Federico Santa Mar\'{\i}a\\
Casilla 110-V, Valpara\'{\i}so, Chile.\\
Phone: +56(32)654-506\\
Fax: +56(32)797-656\\
\vfill
\noindent
PACS: 02.40.Gh, 03.65.Ca, .\\
Keywords: Noncommutative quantum mechanics, gauge symmetry, Landau
problem.


\newpage

\parskip=7pt
\parindent=0pt
\section{Introduction}

A lot of attention has been given recently to the formulation and
possible experimental consequences of extensions of the standard
formalism of quantum mechanics to allow for non-commuting position
operators\cite{Chaichian-et-al,Dayi-Jellal,Banerjee,Gamboa-et-al,Ho-Kao,
Kochan-Demetrian,Mezincescu,Muthukumar-Mitra,Nair-Polychronakos,
Smailagic-Spallucci,Zunger}. Much of this work has been inspired by ideas coming
from the realms of string theory\cite{string-theory} and quantum field theory
(for a review, see \cite{NCQFT-review}).
The standard way of constructing a noncommutative field
theory is to replace the usual product of fields (appearing in the
action in the functional integral) by the so called Moyal or star
product, defined by
\ba
(\phi _1 \star\phi _2) (x) = \exp \left( {i\theta _{\mu \nu } \partial
_x^\mu  \partial _y^\nu  } \right)\phi _1 (x)\left. {\phi _2 (y)}
\right|_{x = y},
\ea
where $\theta _{\mu \nu }$ is an antisymmetric constant matrix. As
$\theta _{0 i }\neq 0$ leads to a non-unitary
theory\cite{non-unitary}, only the elements $\theta _{i j
},\;i,j=1,2,3$, are allowed to be non-vanishing. This leads to the
Moyal commutation relations between the spatial coordinate fields
\ba
x_i\star x_j  - x_j\star x_i  = i\theta _{ij}.
\ea
In order not to spoil the isotropy of space
it is mandatory to choose $\theta _{ij}$ proportional to the
constant antisymmetric matrix $\boldsymbol\epsilon$, $\theta
_{ij}=\theta{\bf\epsilon}_{ij}$, where $\theta$, the
noncommutativity parameter, is a constant with dimension of
$(length)^2$ and
\ba
{\boldsymbol\epsilon}  = \left( {\begin{array}{*{20}c}
   0 & { - 1} & 1  \\
   1 & 0 & { - 1}  \\
   { - 1} & 1 & 0  \\
 \end{array} } \right).
\ea
Inspired by this formalism, many authors have considered an
extension of non-relativistic quantum mechanics, usually referred
to in the literature as noncommutative quantum mechanics (NCQM).
This extended theory is formulated in the same terms as the
standard theory (SQM), that is, in terms of the same dynamical
variables represented by operators in a Hilbert space and a state
vector that evolves according to the Schr\"{o}dinger equation,
\ba
i\hbar \frac{d}
{{dt}}\left| \psi  \right\rangle  = H_{nc}  \left| \psi  \right\rangle
\ea
where $H_{nc}$ is the Hamiltonian for a given system in the
noncommutative theory. The crucial difference with the standard
theory is that in the extended theory the operators representing
the {\em position} of a particle, $X_i$, are no longer assumed to commute
among themselves, but instead the following non-canonical
commutation relations are postulated:
\ba\label{commutation-relations-non-canonical}
[X_i ,X_j ] = i\theta _{ij} ,\quad
[X_i ,P_j ] = i\hbar \delta _{ij} ,\quad
[P_i ,P_j ] = 0.
\ea
with $\theta_{ij}=\theta{\bf\epsilon}_{ij}$ as before.
To completely specify a particular NCQM
system it is necessary to define the Hamiltonian $H_{nc}$, which from here on we shall
simply denote by $H_\theta\equiv H_{nc}$. This Hamiltonian
$H_\theta$ is to be chosen such that it reduces to the Hamiltonian
$H$ for the standard theory in the limit $\theta\to 0$. Two
approaches are found in the literature: (a) simply take
$H_\theta=H$, so that the only difference between SQM and NCQM is
the presence of a nonzero $\theta$ in the commutator of the
position operators\cite{Chaichian-et-al,Muthukumar-Mitra,Nair-Polychronakos,Smailagic-Spallucci};
or (b) naively derive the Hamiltonian from the
Moyal analog of the standard Schr\"{o}dinger {\em wave} equation,
namely
\ba\label{schroedinger-moyal-eqn}
i\hbar \frac{\partial } {{\partial t}}\psi ({\bf x},t) =
H({\bf p}=\frac{\hbar}{i}\nabla,{\bf x})\star\psi ({\bf x},t)
\equiv H_\theta\psi ({\bf x},t),
\ea
where $H({\bf p},{\bf x})$ is the same Hamiltonian as in the
standard theory, so that the $\theta$-dependence enters now solely
through the star product in the equation above\cite{Dayi-Jellal,Gamboa-et-al,Mezincescu}.
In\cite{Mezincescu} it has been shown that, for a Hamiltonian of
the type
\ba
H({\bf p},{\bf x}) = \frac{{{\bf p}^2 }}
{{2m}} + V({\bf x}),
\ea
describing a non-relativistic particle moving in a external
potential, the modified Hamiltonian $H_\theta$ can be simply
obtained by a shift in the argument of the potential, namely,
\ba\label{H-theta-from-moyal}
H_\theta   = \frac{{{\bf p}^2 }}
{{2m}} + V(x_i  - \frac{1}
{{2\hbar }}\theta _{ij} p_j ).
\ea
where $\mathbf{x},\mathbf{p}$ are now canonical variables.

\medskip
As we shall show in the next section, these two approaches
actually lead to the same physical theories. However, these
theories have to be taken with suspicion. In fact, claims have
been presented in the recent literature\cite{Ho-Kao} that a
rigorous derivation of noncommutative quantum mechanics from
noncommutative field theory does not lead to the simple Moyal
Schr\"{o}dinger equation (\ref{schroedinger-moyal-eqn}). For instance,
it is found in\cite{Ho-Kao} that, at tree level in NCQED, there
are no noncommutative corrections to the hydrogen atom spectrum,
contradicting therefore the main conclusion of
reference\cite{Chaichian-et-al}.

In this paper we will argue that the theories obtained by the
approach just described are flawed in a more fundamental way: in
general, they do not share the symmetries possessed by their
commutative counterpart. We shall show nevertheless that there is
natural way of making the transition to the noncommutative quantum
theory, without spoiling the symmetries (such as rotational or
gauge invariance) of the particular system being considered.
Although the prescription to obtain the form of the deformed
Hamiltonian $H_\theta$ corresponding to a given $H$ does not
uniquely define $H_\theta$, it can be considered minimal in a
sense discussed later on. We will see that this prescription will
lead to dramatically different consequences than the standard
approach, the most conspicuos one being that the energy spectrum
of the standard and non-commuting theories are identical, so that
experimental differences between both theories are to be found
only at the level of the detailed structure of the energy
eigenstates, as manifested for instance through transition rates
or expectation values involving the physical position operator.

\section{Rotational symmetry in NCQM}
\label{rotational-symmetry}

In reference\cite{Chaichian-et-al} it has been shown that the
phase-space dynamical variables ${X}_i, {P}_j$ can be expressed
linearly in terms of {\em canonically commuting} variables ${x}_i,
{p}_j$ as
\ba\label{basic-transformation}
 X_i  =  x_i  - \frac{\theta }
{{2\hbar }}\varepsilon _{ij}  p_j ,\quad
 P_i  =  p_i,
\ea
with the inverse transformation given by
\ba\label{basic-transformation-inverse}
 x_i  =  X_i  + \frac{\theta }
{{2\hbar }}\varepsilon _{ij}  P_j ,\quad
 p_i  =  P_i.
\ea
We would like to emphasize that throughout this paper the
capitalized symbols $X_i,P_j$ will always denote the physical position
and momentum operators, both in SQM ($\theta=0$) and NCQM ($\theta\neq 0$).
On the other hand, the lowercase
symbols $x_i,p_j$ will denote the canonically commuting auxiliary
variables defined by (\ref{basic-transformation-inverse}). The
momentum operators $p_i$ and $P_i$ can be interchanged, but $x_i$
coincides with the physical position operator $X_i$ only when
$\theta=0$.

From (\ref{basic-transformation}) it is clear that approaches (a)
and (b), described in the introduction, lead to identical
theories, since
\ba\label{wrong-Htheta}
\frac{{p^2 }}
{{2m}} + V(x_i  - \frac{1}
{{2\hbar }}\theta _{ij} p_j ) = \frac{{P^2 }}
{{2m}} + V(X_i ).
\ea
Let us now consider rotational invariance. Due to the
non-commuting nature of the position operators $X_i$, the
canonical operators $L_i  = \varepsilon _{ijk} X_j P_k,\; i=1,2,3,$
no longer satisfy the angular momentum algebra,
$[L_i,L_j]=i\varepsilon _{ijk}L_k$. However, since the operators
$x_i,  p_j$ are canonically conjugated, it is immediate that the
operators
\ba
J_i  = \varepsilon _{ijk} x_j p_k  = L_i  + \frac{\theta }
{{2\hbar }}\left( {P_i P - P^2 } \right),
\ea
where $P\equiv P_1+P_2+P_3$ and $P^2\equiv P_1^2+P_2^2+P_3^2$, do
and therefore form a representation of the algebra of the rotation
group.

Therefore, the NCQM theory defined by a Hamiltonian $H_\theta$
will describe a rotationally invariant system only if each of the
generators $J_i$ commutes with $H_\theta$. This is clearly not the
case if $H_\theta$ is of the type (\ref{wrong-Htheta}), even for a
central potential $V(R^2)$, with $R^2 \equiv X_i X_i$, since
\ba
R^2  &=& \left( { x_i  - \frac{\theta }
{{2\hbar }}\varepsilon _{ij}  p_j } \right)\left( { x_i  - \frac{\theta }
{{2\hbar }}\varepsilon _{ik}  p_k } \right)
\nn\\
\label{R-squared}
&=& r^2  + \frac{\theta }
{{2\hbar }}\left( {J_1  + J_2  + J_3 } \right) + \frac{{\theta ^2 }}
{{2\hbar ^2 }}\left( {p^2  - p_1 p_2  - p_2 p_3  - p_3 p_1 }
\right),
\ea
and $[R^2,J_i]\neq 0$ for $\theta\neq 0$.

Now, in terms of the auxiliary
canonical variables $x_i,p_j$, the construction of rotationally
invariant operators proceeds as usual. For instance, the
Hamiltonian
\ba\label{H-theta-central-potential}
H_\theta   = \frac{{p^2 }}{{2m}} + V(x^2 )
\ea
clearly commutes with each $J_i$ and reduces to
\ba\label{H-standard-central-potential}
H  = \frac{{P^2 }}{{2m}} + V(X^2 )
\ea
in the limit $\theta\to 0$. We see that the spectra of both
Hamitonians are identical, since these are completely determined
by the commutation relations between the position and momentum
variables. Does this mean that both theories, on the one hand
the standard theory based on the Hamiltonian
(\ref{H-standard-central-potential}), and on the other the
deformed theory based on the non-canonical commutation relations
(\ref{commutation-relations-non-canonical}) and the Hamiltonian
(\ref{H-theta-central-potential}), are unitarily equivalent?
It does not, since the coordinate transformation
\ba
S^{-1}   X_i S =  x_i ,\quad
S^{-1}   P_i S =  p_i
\ea
is not unitary, as it does not preserve the commutators
(\ref{commutation-relations-non-canonical}) which evaluate to
c-numbers.

Clearly the Hamiltonian (\ref{H-theta-central-potential}) is not
the most general rotationally invariant operator reducing to the
standard SQM Hamiltonian (\ref{H-standard-central-potential}) in
the limit $\theta\to 0$. But it can certainly considered to be
``minimal'' in its class.

\medskip

As a particular example we consider the NCQM of the isotropic
three-di\-men\-sion\-al harmonic oscillator, whose dynamics we postulate
to be governed by the rotationally invariant Hamiltonian
\ba
H_\theta   = \frac{{p^2 }} {{2m}} + \frac{1} {2}m\omega ^2 r^2.
\ea
The well-known spectrum of $H_\theta$,
%
\ba
E = \hbar \omega (n_1  + n_2  + n_3  + 3/2),
\ea
has no dependence on $\theta$ and is identical to the commutative
case. We note that many
authors\cite{Kochan-Demetrian,Muthukumar-Mitra,Nair-Polychronakos,Smailagic-Spallucci},
have obtained $\theta$-dependent spectra for this system; this is not
contradictory with our result: it merely reflects the fact that
these authors assume $H_\theta(X,P)=H_0(X,P)$, where $H_0$ is exactly
the same as the Hamiltonian defining the commutative system. This
choice, however, does not lead to a rotationally invariant theory,
as we have pointed out.

$\theta$-dependent differences between both theories will
arise when we consider, for instance, the expectation value of
$ R^2$, which measures the spatial width of the quantum state.
With $R^2$ given by (\ref{R-squared}) and using, for the standard
harmonic oscillator,
\ba
\left\langle {r^2 } \right\rangle  = \frac{E}
{{m\omega ^2 }},\quad
\left\langle {p^2 } \right\rangle  = mE,
\ea
together with $\left\langle {L_i } \right\rangle = \left\langle
{p_i } \right\rangle = 0$ and $\left\langle {p_i p_j}
\right\rangle = 0$ for $i\neq j$, we find for the expectation
value of $R^2$ in a energy eigenstate of energy $E$
\ba
\left\langle {R^2 } \right\rangle  = \frac{E}
{{m\omega ^2 }}\left( {1 + \frac{{m^2 \omega ^2 \theta ^2 }}
{{2\hbar ^2 }}} \right).
\ea
This result can also be written as
\ba
\left\langle {R^2 } \right\rangle  = \left. {\left\langle {R^2 }
\right\rangle } \right|_{\theta  = 0} \left[ {1 + \frac{1}
{2}\left( {\theta /l_\omega ^2 } \right)^2 } \right],
\ea
where $l_\omega\equiv \sqrt{\hbar/m\omega}$ is the oscillator
length of the standard harmonic oscillator.

\section{The Landau electron in SQM}
\label{sqm-landau-electron}

In this section we review the standard quantum mechanical
treatment of a non-relativistic electron moving in the background
of a \emph{uniform} external magnetic field, with special emphasis on the
symmetries of the system. In SQM, such a system is
described by the Hamiltonian
\ba\label{H-landau}
H = \frac{1} {{2m}}\left( {{\mathbf{P}} -
e{\mathbf{A}}({\mathbf{R}})} \right)^2,
\ea
where the physical variables $\mathbf{R},\mathbf{P}$ are
canonically commuting (\ie $\theta=0$) and
${\mathbf{A}}({\mathbf{r}})$ is some vector potential describing
the uniform magnetic field, which we take to point in the
$z$-direction, and we have for simplicity omitted Pauli's spin
term, since it is unchanged in the transition to the
noncommutative theory.

Perhaps the most fundamental property of the Hamiltonian
(\ref{H-landau}) is that the resulting theory is gauge
invariant, that is, the
physical consequences of the theory are unchanged by a gauge
transformation of the vector potential,
\ba
{\mathbf{A}}({\mathbf{r}}) \to {\mathbf{A}}'({\mathbf{r}}) =
{\mathbf{A}}({\mathbf{r}}) + \nabla \chi ({\mathbf{r}}).
\ea
(Here and throughout this section lowercase position variables
such as $\mathbf{r}, x, y$, etc., will denote classical c-number
variables and not operators.)
This invariance can be explicitly represented in terms of a
unitary operator
\ba
T_\chi   = \exp \left[ {i\frac{e}
{\hbar }\chi \left( {\mathbf{R}} \right)} \right],
\ea
under which the fundamental dynamical variables of the theory
transform as
\ba\label{unitary-gauge-transformation}
T_\chi ^\dag  {\mathbf{R}}T_\chi &=& {\mathbf{R}}\\
T_\chi ^\dag  {\mathbf{P}}T_\chi &=& {\mathbf{P}}
+ e\nabla \chi ({\mathbf{R}}),
\ea
so that
\ba\label{H-gauge-transformation}
T_\chi ^\dag  H\left[ {\mathbf{A}} \right]T_\chi   = H\left[
{{\mathbf{A}}'} \right].
\ea
Since the operator $T_\chi$ is unitary, the transformation
property (\ref{H-gauge-transformation}) implies that
the energy spectrum of the theory is gauge invariant.

Gauge invariance also plays a crucial role in establishing that
the theory described by the Landau Hamiltonian (\ref{H-landau})
has the space symmetries of the background magnetic field, that
is, translational symmetry and rotational symmetry around the
magnetic field direction. Let us consider first translational
symmetry. This symmetry is apparently broken once we make a
particular gauge choice for the vector potential, for instance,
the symmetric gauge,
\ba\label{symmetric-gauge}
{\mathbf{A}}({\mathbf{r}}) =  - \frac{1}
{2}{\mathbf{r}} \times {\mathbf{B}} = \frac{B}
{2}( - y,x,0),
\ea
or Landau gauge,
\ba\label{landau-gauge}
{\mathbf{A}}({\mathbf{r}}) = B( - y,0,0).
\ea
Clearly, the Hamiltonian (\ref{H-landau}) no longer commutes with
all three generators of infinitesimal translations, $P_x, P_y,
P_z$. In fact,
\ba
e^{ - i{\mathbf{P}} \cdot {\mathbf{a}}/\hbar } H\left[
{{\mathbf{A}}({\mathbf{r}})} \right]e^{i{\mathbf{P}} \cdot
{\mathbf{a}}/\hbar }  = H\left[ {{\mathbf{A}}({\mathbf{r}} +
{\mathbf{a}})} \right].
\ea
However, in view of the gauge invariance
(\ref{H-gauge-transformation}), it is not actually necessary that
the Hamiltonian $H$ be unchanged under the translation
${\mathbf{r}}\to{\mathbf{r}}+{\mathbf{a}}$; it is enough that the
potential ${\mathbf{A}}({\mathbf{r}} +{\mathbf{a}})$ be a gauge
transform of the original potential ${\mathbf{A}}({\mathbf{r}})$.
But this is just the case if ${\mathbf{B}} = \nabla  \times
{\mathbf{A}}$ is a uniform magnetic field, as can be seen by
performing a Taylor expansion around ${\mathbf{r}}$,
\ba
A_i ({\mathbf{r}} + {\mathbf{a}}) = A_i ({\mathbf{r}}) + \frac{{\partial A_i }}
{{\partial x_j }}({\mathbf{r}})a_j  + \frac{1}
{2}\frac{{\partial ^2 A_i }}
{{\partial x_k \partial x_j }}({\mathbf{r}})a_k a_j  +  \ldots,
\ea
and using
\ba
\frac{{\partial A_i }}
{{\partial x_j }} = \frac{{\partial A_j }}
{{\partial x_i }} - \varepsilon _{ijk} B_k,
\ea
so that
\ba
\frac{{\partial ^{n + 1} A_i }}
{{\partial x_{k_n }  \cdots \partial x_{k_1 } \partial x_j }} = \frac{\partial }
{{\partial x_i }}\frac{{\partial ^n A_j }}
{{\partial x_{k_n }  \cdots \partial x_{k_1 } }},
\ea
we indeed find that
\ba
{\mathbf{A}}({\mathbf{r}} + {\mathbf{a}}) =
{\mathbf{A}}({\mathbf{r}}) + \nabla \chi _{\mathbf{a}}
({\mathbf{r}}),
\ea
where
\ba
\chi _{\mathbf{a}} ({\mathbf{r}}) =  {\mathbf{a}} \cdot
{\mathbf{r}} \times {\mathbf{B}} + {\mathbf{a}} \cdot {\mathbf{A}}
+ \frac{1} {{2!}}a_k \frac{{\partial ({\mathbf{a}} \cdot
{\mathbf{A}})}} {{\partial x_k }} + \frac{1} {{3!}}a_k a_l
\frac{{\partial ({\mathbf{a}} \cdot {\mathbf{A}})}} {{\partial x_k
\partial x_l }} +  \cdots.
\ea
An alternative way of looking at the same problem is to actually
enforce the invariance of the Hamiltonian, but under a set of
generalized translation generators, given by
\ba
{\mathbf{P}_A} = {\mathbf{P}} + e\left( {{\mathbf{A}}({\mathbf{R}})
+ {\mathbf{R}} \times {\mathbf{B}}} \right),
\ea
which also represent the algebra of the Euclidean translation
group,
\ba
[P_{Ai},P_{Aj}] = 0.
\ea
We note that under the gauge transformation
(\ref{unitary-gauge-transformation}) the generalized translation
generators transform covariantly, that is,
\ba
T_\chi ^\dag  {\mathbf{P}}_A T_\chi   = {\mathbf{P}}_{A'},
\ea
where ${\mathbf{A}'}={\mathbf{A}}+\nabla\chi$.

The latter point of view can be also adopted in considering
rotational invariance. We define a set of generalized rotation
generators,
\ba
{\mathbf{L}_A} = {\mathbf{L} + e\mathbf{R}} \times
\left( {{\mathbf{A}}({\mathbf{R}}) + \frac{1}
{2}{\mathbf{R}} \times {\mathbf{B}}} \right)
\ea
which transform covariantly
under gauge transformations and
satisfy the canonical commutation relations
\ba
[L_{Ai},L_{Aj}] = i\hbar \varepsilon _{ijk} L_{Ak},
\ea
provided the magnetic field ${\mathbf{B}}$ is uniform.
Then  the Hamiltonian (\ref{H-landau}) describing the motion of an
electron in a uniform magnetic field $\mathbf{B}=B\hat{\mathbf{z}}$
commutes with $L_{Az}$,
\ba
[H,L_{Az}] = 0.
\ea
This can be explicitly demonstrated, for instance,
in the symmetric gauge (\ref{symmetric-gauge}), where
${\mathbf{L}}_A$ reduces to the canonical
${\mathbf{L}}={\mathbf{R}}\times{\mathbf{P}}$, and $L_z$ indeed
commutes with the Hamiltonian, since in this gauge one has
\ba
H = \frac{1}
{{2m}}\left[ {{\mathbf{P}}^2  + \left( {\frac{{eB}}
{2}} \right)^2 {\mathbf{R}}^2  - eBL_z } \right].
\ea

\section{The Landau electron in NCQM}
\label{ncqm-landau-electron}

All the symmetries described above for the motion of an electron
in the background of a uniform magnetic field will be preserved in
the noncommutative theory if we choose the Hamiltonian of the same
form as in the standard theory, but with the physical variables
${\mathbf{R}}$ and ${\mathbf{P}}$ replaced by the auxiliary
canonical variables ${\mathbf{r}}$ and ${\mathbf{p}}$:
\ba\label{H-theta-landau}
H_\theta  & =& \frac{1}
{{2m}}\left( {{\mathbf{p}} - e{\mathbf{A}}({\mathbf{r}})}
\right)^2\\
&=&
\frac{1}
{{2m}}\left( {{\mathbf{P}} - e{\mathbf{A}}\left( {{\mathbf{R}} + \frac{\theta }
{{2\hbar }}{\mathbf{\tilde P}}} \right)} \right)^2,
\ea
where we have used the customary notation $\tilde P_i\equiv
\varepsilon_{ij}P_j$.
The spectrum of $H_\theta$ is exactly the same as that of the
original Hamiltonian (\ref{H-landau}) defining the standard
theory, that is, the well known Landau level spectrum
\ba
E(p_z ,n) = \frac{{p_z^2 }}
{{2m}} + \hbar \omega _B \left( {n + \frac{1}
{2}} \right),
\ea
where $\omega_B=eB/m$ is the electron cyclotron frequency.
However, as in the case of the harmonic oscillator studied in
section \ref{rotational-symmetry}, the precise shape of the
stationary quantum states is different in both theories, a fact
that is exemplified, for instance, by computing the radius of the
orbit corresponding to a given Landau level. To show this, we will
use the formalism developed in\cite{Ezawa-book} to compute the
expectation value of $\rho_x^2+\rho_y^2$, where
${\boldsymbol{\rho}}$ is the operator representing the position of
the electron with respect to the guiding center $\xi_x, \xi_y$,
which is the quantum analog of the center of the classical
electron orbit. The guiding center coordinates, $\xi_x, \xi_y$,
have the property $[H_\theta,\xi_x]=[H_\theta,\xi_y]=0$, and in
the symmetric gauge (\ref{symmetric-gauge}) they are given by
\ba
\xi _x  = \frac{1}
{2}x + \frac{{\ell _B^2 }}
{\hbar }p_y, \qquad \xi _y  = \frac{1}
{2}y - \frac{{\lbsq}}
{\hbar }p_x,
\ea
where $\lbsq\equiv\hbar/eB$. The position of the electron with respect
to the guiding center is thus given by
\ba
\rho _x  = X - \xi _x & = &\frac{1}
{2}x - \frac{\lbsq}
{\hbar }\left( {1 - \frac{\theta }
{{2\lbsq}}} \right)p_y
\\
\rho _y  = Y - \xi _y & = &\frac{1}
{2}y + \frac{\lbsq}
{\hbar }\left( {1 - \frac{\theta }
{{2\lbsq}}} \right)p_x
\ea
where we have set $p_z=0$ for simplicity. As shown in\cite{Ezawa-book},
the canonical operators $x,y,p_x,p_y$ can be
written in terms of two pairs of independent harmonic oscillator
operators, $(a,a^\dag)$ and $(b, b^\dag)$, so that $\rho_x,
\rho_y$ take the form
\ba
\rho _x & = &\frac{\lb}
{{\sqrt 2 }}\left( {1 - \frac{\theta }
{{4\lbsq}}} \right)i(a - a^\dag  ) + \frac{\theta }
{{4\sqrt 2 \lb}}(b + b^\dag  )
\\
\rho _y & = &\frac{\lb}
{{\sqrt 2 }}\left( {1 - \frac{\theta }
{{4\lbsq}}} \right)(a + a^\dag  ) + \frac{\theta }
{{4\sqrt 2 \lb}}i(b - b^\dag  )
\ea
and hence
\begin{multline}
\rho _x^2  + \rho _y^2  = \lbsq\left( {1 - \frac{\theta }
{{4\lbsq}}} \right)^2 (1 + 2a^\dag  a) + \frac{{\theta ^2 }}
{{16\lbsq}}(1 + 2b^\dag  b) \\
+ \frac{\theta }{2}\left( {1 - \frac{\theta }
{{4\lbsq}}} \right)i(ab - a^\dag  b^\dag  ).
\end{multline}
In terms of the operators $a, a^\dag, b, b^\dag$ the Hamiltonian
and the angular momentum operator $l_z$ read
\ba
H = \hbar \omega _B \left( {a^\dag  a + \frac{1}
{2}} \right)\quad\text{and}\quad
l_z  = \hbar \left( {b^\dag  b - a^\dag  a} \right),
\ea
so that in a state with definite numbers of $a$- and $b$-quanta,
$n$ and $n'=n+l$ respectively, such that the energy is $E=\hbar
\omega_B (n+1/2)$ and the angular momentum is $l_z=\hbar l$, we
have
\ba
\left\langle {\rho _x^2  + \rho _y^2 } \right\rangle _{n,l}  =
\lbsq (2n + 1)\left[ {1 - \frac{\theta } {{2\lbsq}} + \frac{{\theta ^2
}} {{8\lb^4 }}\left( {1 + \frac{l} {{2n + 1}}} \right)} \right]
\ea
%


\section{Gauge invariance in NCQM}
\label{ncqm-gauge-invariance}

The arguments of section (\ref{sqm-landau-electron}) can be
repeated line by line to show that
the Hamiltonian $H_\theta$ (\ref{H-theta-landau}) enjoys the
property
\ba
T_{\theta \chi }^\dag  H_\theta  \left[ {\mathbf{A}} \right]T_{\theta \chi }
= H_\theta  \left[ {{\mathbf{A}}'} \right],
\ea
where ${\mathbf{A}'}={\mathbf{A}}+\nabla\chi$ as before, but now
\ba
T_{\theta \chi }  = \exp \left[ {i\frac{e}
{\hbar }\chi \left( {\mathbf{r}} \right)} \right] = \exp \left[ {i\frac{e}
{\hbar }\chi \left( {{\mathbf{R}} + \frac{\theta }
{{2\hbar }}{\mathbf{\tilde P}}} \right)} \right].
\ea
In terms of the physical variables ${\mathbf{R}}$ and
${\mathbf{P}}$, the corresponding symmetry operation can be
written as
\ba
T_{\theta \chi }^\dag  X_i T_{\theta \chi }
&=& X_i  - \frac{{e\theta }}{{2\hbar }}\varepsilon _{ij}
\partial _j \chi \left( {{\mathbf{R}} + \frac{\theta }
{{2\hbar }}{\mathbf{\tilde P}}} \right)
\\
T_{\theta \chi }^\dag  P_i T_{\theta \chi }
&=& P_i  + e\partial _i \chi \left( {{\mathbf{R}} + \frac{\theta }
{{2\hbar }}{\mathbf{\tilde P}}} \right).
\ea
It is clear that this transformation reduces to the standard one
(\ref{unitary-gauge-transformation}) in the limit $\theta\to 0$.
We note in passing that it is impossible to maintain the standard
transformation (\ref{unitary-gauge-transformation}) in the
noncommutative theory, since the canonical commutator
$[X_i,P_j]=i\hbar\delta_{ij}$ would not be preserved:
\ba
i\hbar \delta _{ij}  = T^\dag  [X_i ,P_j ]T = [X_i ,P_j  +
e\partial _j \chi ({\mathbf{R}})] = i\hbar \delta _{ij}  + e[X_i
,\partial _j \chi ({\mathbf{R}})],
\ea
where the last term is non-vanishing in NCQM.

%
%

\bigskip
\noindent{\Large \bf Acknowledgments}
\medskip
\ret This work was supported by CONICYT (Chile) under Grants
Fondecyt PLC-8000017 and 1000710, and by a C\'{a}tedra Presidencial.
P.G. would like to thank I.~Schmidt for his support.

\newcommand{\prd}[3]{Phys.~Rev.~{\bf D#1}, #3 (#2)}
\newcommand{\prl}[3]{Phys.~Rev.~Lett.~{\bf #1}, #3 (#2)}
\newcommand{\plb}[3]{Phys.~Lett.~{\bf B#1}, #3 (#2)}
\newcommand{\pla}[3]{Phys.~Lett.~{\bf A#1}, #3 (#2)}
\newcommand{\hepth}[1]{hep-th/#1}
\newcommand{\hepph}[1]{hep-ph/#1}
\newcommand{\condmat}[1]{cond-mat/#1}
\newcommand{\mpla}[3]{Mod.~Phys.~Lett.~{\bf A#1}, #3 (#2)}
\newcommand{\jhep}[3]{JHEP~{\bf #1}, #3 (#2)}
\newcommand{\rmp}[3]{Rev.~Mod.~Phys.~{\bf #1}, #3 (#2)}
\newcommand{\npb}[3]{Nucl.~Phys.~{\bf B#1}, #3 (#2)}
\newcommand{\epjc}[3]{Eur.~Phys.~J.~{\bf C#1}, #3 (#2)}


\begin{thebibliography}{99}
\bibitem{string-theory}
N.~Seiberg and E.~Witten, \jhep{09}{1999}{032};
A.~Connes, M.R.~Douglas and A.S.~Schwarz, \jhep{02}{1998}{003}.

\bibitem{NCQFT-review}
M.R.~Douglas and N.A.~Nekrasov, \rmp{73}{2002}{977};
I.~Hinchliffe and N.~Kersting, \hepph{0205040}.

\bibitem{non-unitary}
J.~Gomis and T.~Mehen, \npb{591}{2000}{265};
M.~Chaichian, A.~Demichev, P.~Presnajder and A.~Tureanu,
\epjc{20}{2001}{767}.

\bibitem{Smailagic-Spallucci}
A.Smailagic and E.Spallucci, \prd{65}{2002}{107701}.

\bibitem{Ho-Kao}
P-M.~Ho and H-C.~Kao, \prl{88}{2002}{151602}.

\bibitem{Banerjee}
R.~Banerjee, \hepth{0106280}.

\bibitem{Dayi-Jellal}
\"{O}.~Dayi and A.~Jellal, \pla{287}{2001}{349}.

\bibitem{Gamboa-et-al}
J.~Gamboa, F.~M\'{e}ndez, M.~Loewe and J.C.~Rojas,
\mpla{16}{2001}{2075}.

\bibitem{Muthukumar-Mitra}
B.~Muthukumar and P.~Mitra, \hepth{0204149}.

\bibitem{Mezincescu}
L.~Mezincescu, \hepth{0007046}.

\bibitem{Nair-Polychronakos}
V.P.~Nair and A.P.~Polychronakos, \plb{505}{2001}{267}

\bibitem{Kochan-Demetrian}
D.~Kochan and M.~Demetrian, \hepth{0102050}.

\bibitem{Zunger}
Y.~Zunger, JHEP 0104, 039 (2001).

\bibitem{Chaichian-et-al}
M.~Chaichian, M.M.~Sheikh-Jabbari and A.~Tureanu,
\prl{86}{2001}{2716}.

\bibitem{Ezawa-book}
Z.~F.~Ezawa, {\em Quantum Hall Effects}, World Scientific,
Singapore (2000).

\end{thebibliography}
\end{document}